\documentclass[a4paper,11pt]{article}
\usepackage{pos}

\def \Nc {{N_{c}}}

\title{Testing importance sampling on a quantum annealer for strong coupling $SU(3)$ gauge theory
}

\author*[a]{Jangho Kim}
\author[a,b]{Thomas Luu}
\author[c]{Wolfgang Unger}

\affiliation[a]{Institute for Advanced Simulation (IAS-4) \& JARA-HPC, Forschungszentrum J{\"u}lich, Wilhelm-Johnen-Stra{\ss}e, 52428, J{\"u}lich, Germany}

\affiliation[b]{Institut f{\"u}r Kernphysik (IKP-3), Forschungszentrum J{\"u}lich, Wilhelm-Johnen-Stra{\ss}e, 52428, J{\"u}lich, Germany}
\affiliation[c]{Fakult{\"a}t f{\"u}r Physik, Bielefeld University, Universit{\"a}tsstra{\ss}e 25, 33615 Bielefeld, Germany}
\emailAdd{j.kim@fz-juelich.de}
\emailAdd{t.luu@fz-juelich.de}
\emailAdd{wunger@physik.uni-bielefeld.de}

\abstract{
$SU(N_c)$ gauge theories in the strong coupling limit~\cite{Prokofev2001,Adams:2003cca,Fromm,deForcrand:2009dh,Kim2016} can be described by integer variables representing monomers, dimers and baryon loops. We demonstrate how the D-wave quantum annealer~\cite{PhysRevB.81.134510,PhysRevB.82.024511} can perform importance sampling on $U(N_c)$ gauge theory in the strong coupling formulation of this theory~\cite{Kim:2023sie}. In addition to causing a sign problem~\cite{deForcrand2010,Gattringer2016b} in importance sampling, baryon loops induce a complex QUBO matrix which cannot be optimized by the D-Wave annealer. Instead we show that simulating the sign-problem free quenched action on the D-Wave is sufficient when combined with a sign reweighting method.
As the first test on $SU(3)$ gauge theory, we simulate on $2 \times 2$ lattice and compare the results with its analytic solutions.
}

\FullConference{The 40th International Symposium on Lattice Field Theory (Lattice 2023)\\
July 31st - August 4th, 2023\\
Fermi National Accelerator Laboratory\\}

\begin{document}
\maketitle

\section{Introduction}
The $SU(N_c)$ gauge theory in the strong coupling limit has interesting features like the nuclear liquid-gas transition~\cite{Kim2023}.
Simulations are typically performed using the worm algorithm on a classical computer. However, in the low temperature region the worm algorithm is notoriously inefficient. This is unfortunate since investigations of the first order nuclear transition with heavy quark masses requires simulations exactly in this regime. Since simulating on a quantum annealer does not depend on the temperature, we develop an updating method on the D-wave quantum computer using the QUBO formalism.
While the worm algorithm updates one link or one site per worm step, which scales inefficiently, the update algorithm using a quantum annealer can update $2\times2$, $4\times4$ or $2^4$ sub-volumes essentially all at once since the quantum annealer can sample important configurations in a few microsecond or less.
This feature of the quantum annealer potentially provides a means for performing global updates on larger volumes, an algorithm of which we are currently developing and investigating. Here we present the first results of $SU(3)$ on $2 \times 2$ volume with periodic boundary condition.

\section{QUBO matrix for $SU(N_c)$ gauge theory in the strong coupling limit}
We have recently demonstrated that importance sampling for $U(N_c)$ gauge theory in the strong coupling limit is feasible on D-wave quantum annealer~\cite{Kim:2023sie}.
Here we extend our study to $SU(N_c)$ by including baryon loops.
The partition function of $SU(N_c)$ gauge theory using dual variables is
\begin{align}
    Z=\sum_{\{k,n,\ell\}}
\underbrace{\prod_{b=(x,\hat{\mu})}\frac{(N_c-k_b)!}{N_c!k_b!}\gamma^{2k_b\delta_{\hat{0},\hat{\mu}}}}_{\text{meson hops}}
\underbrace{\prod_{x}\frac{N_c!}{n_x!}(2am_q)^{n_x}}_{\text{chiral condensate}}
\underbrace{\prod_{\ell}B(\ell,\mu)}_{\text{baryon hops}}\,,
\end{align}
where 
\begin{align}
B(\ell,\mu)=\dfrac{1}{\prod_{x\in\ell}N_c!}\sigma(\ell)\gamma^{\Nc N_{\hat{0}}(\ell)} \exp{(\Nc N_t \omega_{\ell}a_t\mu)}\,,
\quad \sigma(\ell)=(-1)^{\omega_{\ell}+N_{-}(\ell)+1}\prod_{b=(x,\hat{\mu})\in\ell}\eta_{\hat{\mu}}(x)\,.
\end{align}
Here, $N_{-}(\ell)$ is the negative direction baryon segments and $\eta_{\hat{\mu}}(x)$ is the staggered phase.

\begin{figure}[h]
    \centering
    \includegraphics[width=0.4\textwidth]{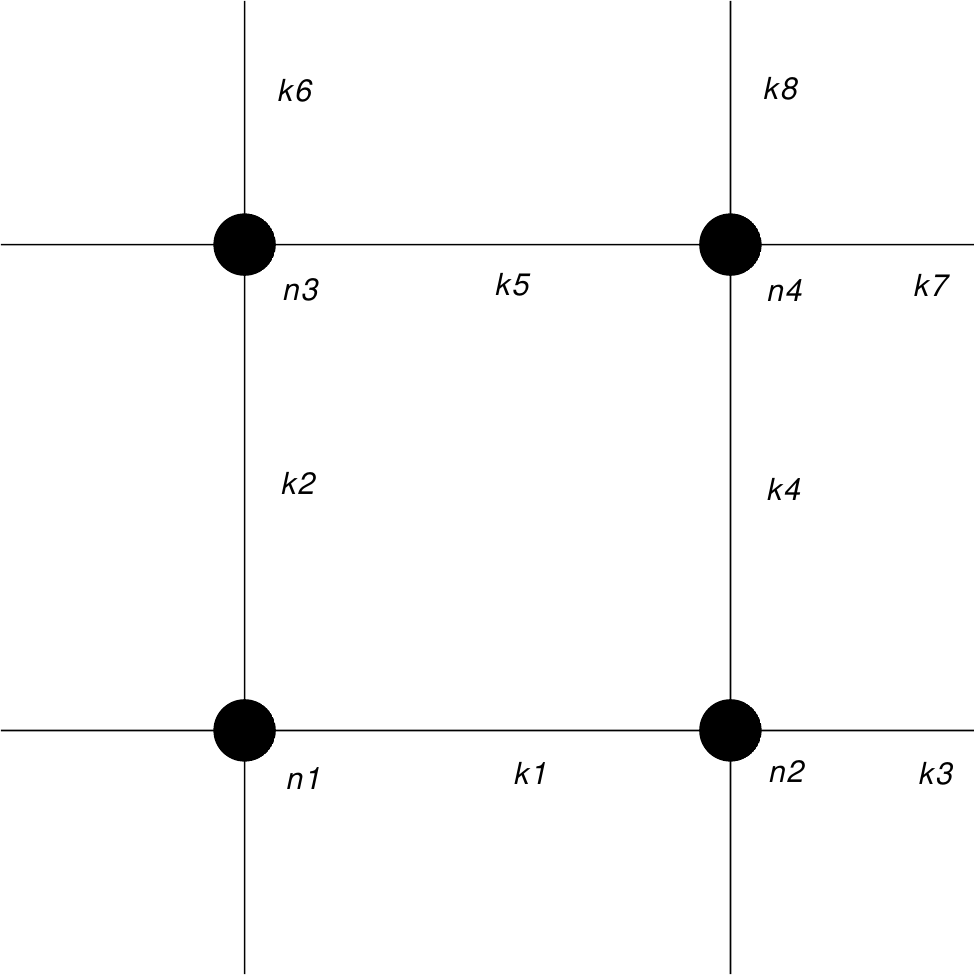}
    \caption{The coordinate convention used in this proceedings.}
    \label{fig:2x2}
\end{figure}
\noindent We use the coordinate convention shown in Fig.~\ref{fig:2x2} where  the specific example of the $2 \times 2$ lattice is given.
\begin{table}[hp]
    \centering
    \footnotesize{
    \begin{tabular}{c|ccc|r| r| c }
         Index & $\sum_{\ell}N_{\hat{0}}(\ell)$ & $\sum_{\ell}\omega_{\ell}$ & $\sum_{\ell}\sigma$ & Baryon configuration & Occupied sites & Solution vector \\ 
         \hline\hline
1  &  0  & 0  & 1    & ( 0,  0,  0,  0,  0,  0,  0,  0) & ()              & (0, 0, 0, 0, 0, 0, 0, 0, 0) \\
\hline
2  &  0  & 0  & -1   & ( 1,  0,  1,  0,  0,  0,  0,  0) & ($n_1$,$n_2$)            & (1, 0, 0, 0, 0, 0, 0, 0, 0) \\
3  &  0  & 0  & -1   & (-1,  0, -1,  0,  0,  0,  0,  0) & ($n_1$,$n_2$)            & (1, 0, 0, 0, 0, 0, 0, 0, 0) \\
\hline
4  &  0  & 0  & -1   & ( 0,  0,  0,  0,  1,  0,  1,  0) & ($n_3$,$n_4$)            & (0, 1, 0, 0, 0, 0, 0, 0, 0) \\
5  &  0  & 0  & -1   & ( 0,  0,  0,  0, -1,  0, -1,  0) & ($n_3$,$n_4$)            & (0, 1, 0, 0, 0, 0, 0, 0, 0) \\
\hline
6  &  2  & 1  & 1    & ( 0,  1,  0,  0,  0,  1,  0,  0) & ($n_1$,$n_3$)            & (0, 0, 1, 0, 0, 0, 0, 0, 0) \\
\hline
7  &  2  & -1 &  1   & ( 0, -1,  0,  0,  0, -1,  0,  0) & ($n_1$,$n_3$)            & (0, 0, 0, 1, 0, 0, 0, 0, 0) \\
\hline
8  &  2  & 1  & 1    & ( 0,  0,  0,  1,  0,  0,  0,  1) & ($n_2$,$n_4$)            & (0, 0, 0, 0, 1, 0, 0, 0, 0) \\
\hline
9  &  2  & -1 &  1   & ( 0,  0,  0, -1,  0,  0,  0, -1) & ($n_2$,$n_4$)            & (0, 0, 0, 0, 0, 1, 0, 0, 0) \\
\hline
10 &  0  & 0  & 1    & ( 1,  0,  1,  0,  1,  0,  1,  0) & ($n_1$,$n_2$,$n_3$,$n_4$)      & (1, 1, 0, 0, 0, 0, 0, 0, 0) \\
11 &  0  & 0  & 1    & (-1,  0, -1,  0,  1,  0,  1,  0) & ($n_1$,$n_2$,$n_3$,$n_4$)      & (1, 1, 0, 0, 0, 0, 0, 0, 0) \\
12 &  0  & 0  & 1    & ( 1,  0,  1,  0, -1,  0, -1,  0) & ($n_1$,$n_2$,$n_3$,$n_4$)      & (1, 1, 0, 0, 0, 0, 0, 0, 0) \\
13 &  0  & 0  & 1    & (-1,  0, -1,  0, -1,  0, -1,  0) & ($n_1$,$n_2$,$n_3$,$n_4$)      & (1, 1, 0, 0, 0, 0, 0, 0, 0) \\
\hline
14 &  4  & 2  & 1    & ( 0,  1,  0,  1,  0,  1,  0,  1) & ($n_1$,$n_2$,$n_3$,$n_4$)      & (0, 0, 1, 0, 1, 0, 0, 0, 0) \\
15 &  4  & -2 &  1   & ( 0, -1,  0, -1,  0, -1,  0, -1) & ($n_1$,$n_2$,$n_3$,$n_4$)      & (0, 0, 0, 1, 0, 1, 0, 0, 0) \\
16 &  4  & 0  & 1    & ( 0,  1,  0, -1,  0,  1,  0, -1) & ($n_1$,$n_2$,$n_3$,$n_4$)      & (0, 0, 1, 0, 0, 1, 0, 0, 0) \\
17 &  4  & 0  & 1    & ( 0, -1,  0,  1,  0, -1,  0,  1) & ($n_1$,$n_2$,$n_3$,$n_4$)      & (0, 0, 0, 1, 1, 0, 0, 0, 0) \\
\hline
18 &  2  & 0  & 1    & ( 0,  0, -1,  0,  0,  1,  1, -1) & ($n_1$,$n_2$,$n_3$,$n_4$)      & (0, 0, 0, 0, 0, 0, 1, 0, 0) \\
19 &  2  & 0  & 1    & ( 0,  0,  1,  0,  0, -1, -1,  1) & ($n_1$,$n_2$,$n_3$,$n_4$)      & (0, 0, 0, 0, 0, 0, 1, 0, 0) \\
20 &  2  & 0  & 1    & ( 0,  1,  1, -1,  0,  0, -1,  0) & ($n_1$,$n_2$,$n_3$,$n_4$)      & (0, 0, 0, 0, 0, 0, 1, 0, 0) \\
21 &  2  & 0  & 1    & ( 0, -1, -1,  1,  0,  0,  1,  0) & ($n_1$,$n_2$,$n_3$,$n_4$)      & (0, 0, 0, 0, 0, 0, 1, 0, 0) \\
22 &  2  & 0  & 1    & ( 1, -1,  0,  1, -1,  0,  0,  0) & ($n_1$,$n_2$,$n_3$,$n_4$)      & (0, 0, 0, 0, 0, 0, 1, 0, 0) \\
23 &  2  & 0  & 1    & (-1,  1,  0, -1,  1,  0,  0,  0) & ($n_1$,$n_2$,$n_3$,$n_4$)      & (0, 0, 0, 0, 0, 0, 1, 0, 0) \\
24 &  2  & 0  & 1    & (-1,  0,  0,  0,  1, -1,  0,  1) & ($n_1$,$n_2$,$n_3$,$n_4$)      & (0, 0, 0, 0, 0, 0, 1, 0, 0) \\
25 &  2  & 0  & 1    & ( 1,  0,  0,  0, -1,  1,  0, -1) & ($n_1$,$n_2$,$n_3$,$n_4$)      & (0, 0, 0, 0, 0, 0, 1, 0, 0) \\
26 &  2  & 0  & -1   & ( 1,  0,  0,  0,  0,  1,  1, -1) & ($n_1$,$n_2$,$n_3$,$n_4$)      & (0, 0, 0, 0, 0, 0, 1, 0, 0) \\
27 &  2  & 0  & -1   & (-1,  0,  0,  0,  0, -1, -1,  1) & ($n_1$,$n_2$,$n_3$,$n_4$)      & (0, 0, 0, 0, 0, 0, 1, 0, 0) \\
28 &  2  & 0  & -1   & (-1,  1,  0, -1,  0,  0, -1,  0) & ($n_1$,$n_2$,$n_3$,$n_4$)      & (0, 0, 0, 0, 0, 0, 1, 0, 0) \\
29 &  2  & 0  & -1   & ( 1, -1,  0,  1,  0,  0,  1,  0) & ($n_1$,$n_2$,$n_3$,$n_4$)      & (0, 0, 0, 0, 0, 0, 1, 0, 0) \\
30 &  2  & 0  & -1   & ( 0, -1, -1,  1, -1,  0,  0,  0) & ($n_1$,$n_2$,$n_3$,$n_4$)      & (0, 0, 0, 0, 0, 0, 1, 0, 0) \\
31 &  2  & 0  & -1   & ( 0,  1,  1, -1,  1,  0,  0,  0) & ($n_1$,$n_2$,$n_3$,$n_4$)      & (0, 0, 0, 0, 0, 0, 1, 0, 0) \\
32 &  2  & 0  & -1   & ( 0,  0,  1,  0,  1, -1,  0,  1) & ($n_1$,$n_2$,$n_3$,$n_4$)      & (0, 0, 0, 0, 0, 0, 1, 0, 0) \\
33 &  2  & 0  & -1   & ( 0,  0, -1,  0, -1,  1,  0, -1) & ($n_1$,$n_2$,$n_3$,$n_4$)      & (0, 0, 0, 0, 0, 0, 1, 0, 0) \\
\hline
34 &  2  & 1  & 1    & ( 0,  1,  1,  0,  0,  0, -1,  1) & ($n_1$,$n_2$,$n_3$,$n_4$)      & (0, 0, 0, 0, 0, 0, 0, 1, 0) \\
35 &  2  & 1  & 1    & ( 0,  0, -1,  1,  0,  1,  1,  0) & ($n_1$,$n_2$,$n_3$,$n_4$)      & (0, 0, 0, 0, 0, 0, 0, 1, 0) \\
36 &  2  & 1  & 1    & ( 1,  0,  0,  1, -1,  1,  0,  0) & ($n_1$,$n_2$,$n_3$,$n_4$)      & (0, 0, 0, 0, 0, 0, 0, 1, 0) \\
37 &  2  & 1  & 1    & (-1,  1,  0,  0,  1,  0,  0,  1) & ($n_1$,$n_2$,$n_3$,$n_4$)      & (0, 0, 0, 0, 0, 0, 0, 1, 0) \\
38 &  2  & 1  & -1   & (-1,  1,  0,  0,  0,  0, -1,  1) & ($n_1$,$n_2$,$n_3$,$n_4$)      & (0, 0, 0, 0, 0, 0, 0, 1, 0) \\
39 &  2  & 1  & -1   & ( 1,  0,  0,  1,  0,  1,  1,  0) & ($n_1$,$n_2$,$n_3$,$n_4$)      & (0, 0, 0, 0, 0, 0, 0, 1, 0) \\
40 &  2  & 1  & -1   & ( 0,  0, -1,  1, -1,  1,  0,  0) & ($n_1$,$n_2$,$n_3$,$n_4$)      & (0, 0, 0, 0, 0, 0, 0, 1, 0) \\
41 &  2  & 1  & -1   & ( 0,  1,  1,  0,  1,  0,  0,  1) & ($n_1$,$n_2$,$n_3$,$n_4$)      & (0, 0, 0, 0, 0, 0, 0, 1, 0) \\
\hline
42 &  2  & -1 &  1   & ( 0, -1, -1,  0,  0,  0,  1, -1) & ($n_1$,$n_2$,$n_3$,$n_4$)      & (0, 0, 0, 0, 0, 0, 0, 0, 1) \\
43 &  2  & -1 &  1   & ( 0,  0,  1, -1,  0, -1, -1,  0) & ($n_1$,$n_2$,$n_3$,$n_4$)      & (0, 0, 0, 0, 0, 0, 0, 0, 1) \\
44 &  2  & -1 &  1   & (-1,  0,  0, -1,  1, -1,  0,  0) & ($n_1$,$n_2$,$n_3$,$n_4$)      & (0, 0, 0, 0, 0, 0, 0, 0, 1) \\
45 &  2  & -1 &  1   & ( 1, -1,  0,  0, -1,  0,  0, -1) & ($n_1$,$n_2$,$n_3$,$n_4$)      & (0, 0, 0, 0, 0, 0, 0, 0, 1) \\
46 &  2  & -1 &  -1  & ( 1, -1,  0,  0,  0,  0,  1, -1) & ($n_1$,$n_2$,$n_3$,$n_4$)      & (0, 0, 0, 0, 0, 0, 0, 0, 1) \\
47 &  2  & -1 &  -1  & (-1,  0,  0, -1,  0, -1, -1,  0) & ($n_1$,$n_2$,$n_3$,$n_4$)      & (0, 0, 0, 0, 0, 0, 0, 0, 1) \\
48 &  2  & -1 &  -1  & ( 0,  0,  1, -1,  1, -1,  0,  0) & ($n_1$,$n_2$,$n_3$,$n_4$)      & (0, 0, 0, 0, 0, 0, 0, 0, 1) \\
49 &  2  & -1 &  -1  & ( 0, -1, -1,  0, -1,  0,  0, -1) & ($n_1$,$n_2$,$n_3$,$n_4$)      & (0, 0, 0, 0, 0, 0, 0, 0, 1) \\
    \end{tabular}
    }
    \caption{All possible baryon configurations on $2 \times 2$ lattice.}
    \label{tab:baryon}
\end{table}

In Table~\ref{tab:baryon} we provide all possible baryon configurations for the $2 \times 2$ volume with periodic boundary conditions. The temporal direction baryon segments($N_{\hat{0}}(\ell)$), winding number($\omega_{\ell}$) and sign($\sigma_{\ell}$) on a baryon loop $\ell$, and occupied sites of each configuration are also provided. 
The sign quenched weight of each configuration is 
\begin{align}
    S_{B}=-\log{[1/(N_c!)^{N_{\text{occ}}}]} - N_c \sum_{\ell}N_{\hat{0}}(\ell) \log{\gamma}-N_c N_{\tau} \sum_{\ell}\omega_{\ell} \mu \frac{a}{\gamma} \,,
\end{align}
where $\gamma=a/a_t$ is the anisotropy~\cite{deForcrand2017} and $N_{\text{occ}}$ is the number of baryon occupied sites.
The configurations (18-33) have exactly the same weight but half of them have positive sign and the other have negative sign. Hence the contribution from these configurations will be cancelled with sufficiently large statistics. The same occurs for the configurations (34-41) and (42-49). These configurations therefore just add noise to our stochastic simulations.
Therefore only configurations (1-17) are taken into account in this simulation and the size of solution vector is reduced from 9 qubits to 6 qubits. As the 3 qubits are reduced, the validity rate is also increased from 0.5\% to 2\%.
Since configurations within (10-13) and (14-17) are combinations of certain two configurations within (2-9), we do not explicitly include them in the construction of the weight matrix that follows.

The action part of QUBO matrix can be written in block form,
\begin{align}\label{eqn:weight matrix}
    W=\left[ \begin{array}{ccc}
        D & 0 & 0 \\
        0 & M & 0 \\
        0 & 0 & B
    \end{array} 
    \right]
\end{align}
The matrices $D$ and $M$ are provided by Eqs.~(26),(28),(30), and in Ref.~\cite{Kim:2023sie}.
The baryon matrix $B$ in Eq.~\ref{eqn:weight matrix} is diagonal with components
\begin{align}
    diag(B)=\left[ \begin{array}{c}
        4\log(N_c!) \\
        4\log(N_c!) \\
        2\log(N_c!)-2N_c\log{\gamma}-N_c\gamma a\mu\\
        2\log(N_c!)-2N_c\log{\gamma}+N_c\gamma a\mu\\
        2\log(N_c!)-2N_c\log{\gamma}-N_c\gamma a\mu\\
        2\log(N_c!)-2N_c\log{\gamma}+N_c\gamma a\mu\\
    \end{array} 
    \right]
\end{align}

The constraint matrix also has additional modifications since baryon loops cannot co-exist with monomers or dimers at the same site or bond in the strong coupling limit. This changes by including gauge corrections, and in this case mesonic and baryonic objects can co-exist~\cite{deForcrand2014, Gagliardi2019,Kim2023}. This scenario, however, is not covered in these proceedings.

\begin{tiny}
\begin{align}
   A \cdot x + b = &\left(\begin{array}{rrrrrrrrrrrrrrrr|rrrrrrrr|rrrrrr}
    2 & 1 & 2 & 1 & 2 & 1 & 0 & 0 & 0 & 0 & 2 & 1 & 0 & 0 & 0 & 0 & 2 & 1 & 0 & 0 & 0 & 0 & 0 & 0 & 3 & 0 & 3 & 0 & 3 & 3\\ 
    2 & 1 & 0 & 0 & 2 & 1 & 2 & 1 & 0 & 0 & 0 & 0 & 0 & 0 & 2 & 1 & 0 & 0 & 2 & 1 & 0 & 0 & 0 & 0 & 3 & 0 & 0 & 3 & 3 & 3\\
    0 & 0 & 2 & 1 & 0 & 0 & 0 & 0 & 2 & 1 & 2 & 1 & 2 & 1 & 0 & 0 & 0 & 0 & 0 & 0 & 2 & 1 & 0 & 0 & 0 & 3 & 3 & 0 & 3 & 3\\
    0 & 0 & 0 & 0 & 0 & 0 & 2 & 1 & 2 & 1 & 0 & 0 & 2 & 1 & 2 & 1 & 0 & 0 & 0 & 0 & 0 & 0 & 2 & 1 & 0 & 3 & 0 & 3 & 3 & 3\\
  \end{array}
  \right)
  \\
  \times
   &\left(  \begin{array}{rrrrr|rrrrr|rrrrrr}
		 k_1^{(1)} & k_1^{(2)} & \cdots & k_8^{(1)}  & k_8^{(2)}  & n_1^{(1)}  & n_1^{(2)}  & \cdots & n_4^{(1)}  & n_4^{(2)}  & b_1 & b_2 &
     b_3 &
     b_4 &
     b_5 &
     b_6 \\
  \end{array}
  \right) ^{T}
  -
  \left(  \begin{array}{r}
     3  \\
     3  \\
     3  \\
     3  \\
  \end{array}
  \right)=0
\end{align}
\end{tiny}

Each row corresponds to the site $n_1, n_2, n_3, n_4$. For example, the first column of baryon part $(3,3,0,0)^{T}$ corresponds to configurations (2-3) in the Table~\ref{tab:baryon} and they occupy the sites $n_1$ and $n_2$.
When if $b_1=1$, for example, monomers at those sites $n_1,n_2$ or dimers at the bonds $k_1,k_2,k_3,k_6$ must be $0$ to satisfy the constraint.

\section{Results}
\begin{figure}[b]
    \centering
    \includegraphics[width=0.49\textwidth]{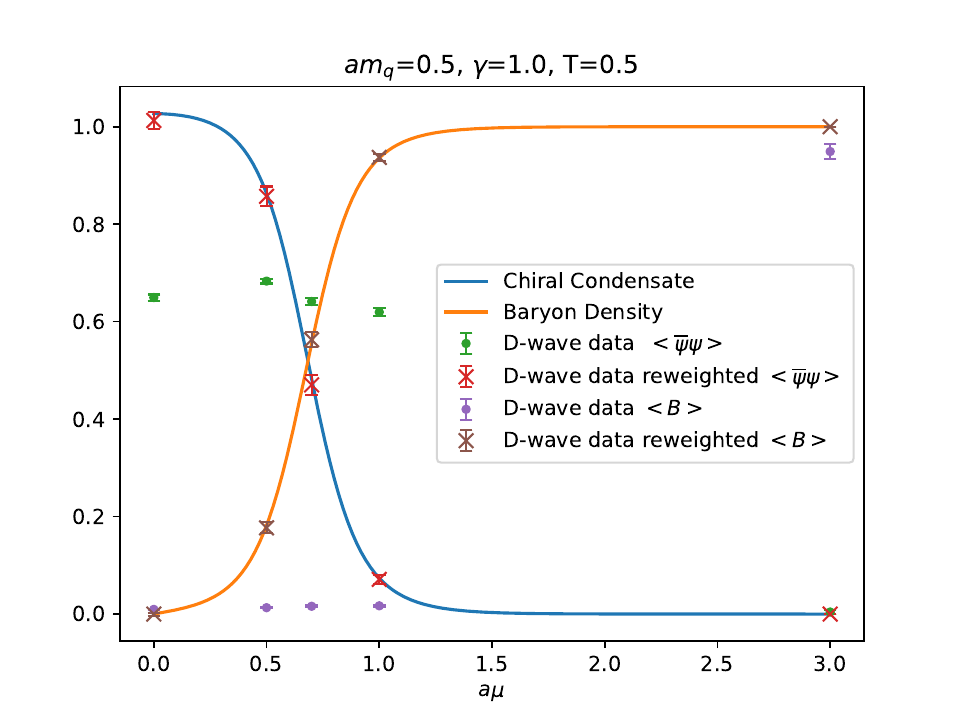}
    \includegraphics[width=0.49\textwidth]{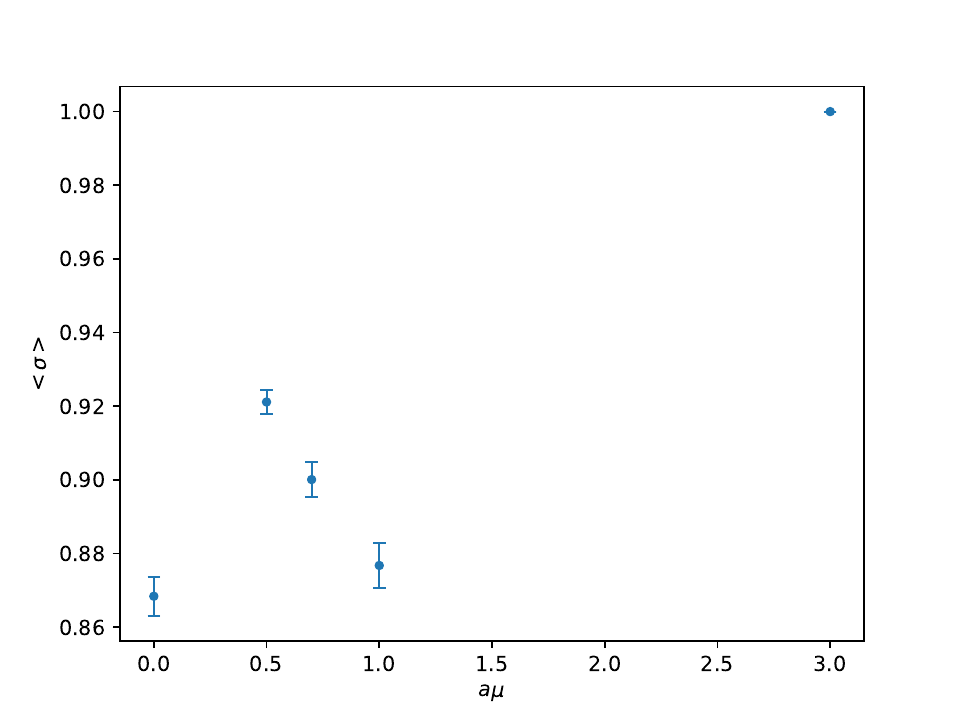}
    \caption{Left: chiral condensate (blue) and baryon density (orange) as a function of the chemical potential $a\mu$ at quark mass $am_q=0.5$, and temperature $T=1.0$ on $2 \times 2$ lattice. The circle points are D-wave raw data and the cross points are reweighted data. Right: the average sign problem.}
    \label{fig:results}
\end{figure}

We measure two observables:  the chiral condensates and baryon density.  These quantities are expressed in dual variables as
\begin{align}
    \left\langle \overline{\psi} \psi \right\rangle &= \frac{1}{2am_q V} \left\langle \sum_x n_x \right\rangle \\
    \left\langle B \right\rangle &= \frac{1}{N_s} \left\langle \sum_\ell w_\ell \right\rangle
\end{align}

The presence of baryon loops induces a sign problem.  Within the QUBO formalism this would correspond to a QUBO Matrix with complex components.  Unfortunately the quantum annealer cannot accept complexed-valued QUBO matrices at this time.  
To circumvent this deficiency, we simulate instead using the sign-quenched action and apply the sign reweighting method to our observables,
\begin{align}
    \langle O \rangle = \frac{\langle \sigma O \rangle }{\langle \sigma \rangle }
\end{align}
We also applied the reweighting method~\cite{Ferrenberg:1988yz} used in Ref.~\cite{Kim:2023sie} and present the results in Fig.~\ref{fig:results}.

The blue and orange solid lines are the analytic solutions of the chiral condensate and baryon density, respectively. The circle points are the D-wave raw data. The reweighting method is shown by the cross data points.  Here the results are improved drastically agree well with their analytic solutions. The averaged sign presented in Fig.~\ref{fig:results} is a residual sign as (18-49) configurations have already been resummed. By this resummation, the sign problem becomes very mild.

\section{Conclusion}

We present the results of $SU(3)$ gauge theory in the strong coupling limit on the $2 \times 2$ lattice.
We find that importance sampling of $SU(3)$ is possible on the D-wave quantum annealer as well as for $U(N_c)$ gauge theory as demonstrated in Ref.~\cite{Kim:2023sie}.
We plan to extend our method to the larger $2^4$ volume with fixed boundary conditions.
Fixed boundary conditions reduces drastically both the size of the solution vector and the number of baryon configurations such that all possible baryon configurations with a given boundary condition can be included by adding only a few qubits to the solution vector. 
For simulating even larger volumes our strategy entails using D-wave to update configurations first on a $2^4$ sub-volume with a specific boundary condition. Both even sites and then odd sites of the sub-volume will be updated, and the process will be repeated on adjacent sub-volumes with boundaries dictated by updates on prior sub-volumes. In this manner we anticipate less correlated data compared to traditional, classical updating schemes.

\begin{acknowledgments}
The authors gratefully acknowledge the J\"ulich Supercomputing Centre (https://www.fzjuelich.de/ias/jsc) for funding this project by providing computing time on the D-Wave Advantage™ System JUPSI through the J\"ulich UNified Infrastructure for Quantum computing (JUNIQ).
J.K. and T. L. were supported by the Deutsche Forschungsgemeinschaft (DFG, German Research Foundation) through the funds provided to the Sino-German Collaborative Research Center TRR110 "Symmetries and the Emergence of Structure in QCD" (DFG Project-ID 196253076 - TRR 110).
W.U. is supported by the Deutsche Forschungsgemeinschaft (DFG) through the CRC-TR 211 'Strong-interaction matter under extreme conditions'– project number 315477589 – TRR 211.
\end{acknowledgments}

\bibliography{references}



\end{document}